\documentclass[aps,floatfix,showpacs]{revtex4}
\usepackage{graphicx}
\usepackage{amsmath,amssymb}
\usepackage[english,activeacute] {babel}

\begin{document}

\title{Constraints on economical 331 models from
mixing of $K$, $B_d$ and $B_s$ neutral mesons} 

\author{J.M.\ Cabarcas$^a$, D.\ G\'omez Dumm$^a$ and R.\ Martinez$^b$}

\affiliation{ $^a$ IFLP, CONICET -- Dpto.\ de F\'{\i}sica, Universidad
Nacional de La Plata, \\ C.C. 67, 1900 La Plata, Argentina. \\
$^b$ Dpto.\ de F\'{\i}sica, Universidad Nacional, Bogot\'a, Colombia.}

\begin{abstract}
We analyze the effect of flavor changing neutral currents within 331
models. In particular, we concentrate in the so-called ``economical''
models, which have a minimal scalar sector. Taking into account the
experimental measurements of observables related to neutral $K$ and $B$
meson mixing, we study the resulting bounds for angles and phases in the
mixing matrix for the down quark sector, and the mass and mixing
parameters related to the new $Z'$ gauge boson.
\end{abstract}

\pacs{11.30.Hv, 12.15.Ff, 12.60.Cn}

\maketitle

\section{Introduction}

In the Standard Model (SM), processes mediated by flavor changing neutral
currents (FCNC) are forbidden at the tree level, occurring only through
diagrams with one or more loops. This is consistent with experimental
observations, which show that the corresponding physical observables
appear to be highly suppressed. Now, it is important to determine if the
experimental results are in fact in agreement with SM predictions, and
to establish what room is available for the presence of new physics.

We analyze here this problem in the framework of the so-called 331 models,
in which the SM gauge symmetry group is enlarged to $SU(3)_C\otimes
SU(3)_L\otimes U(1)_X$~\cite{pleitez,331,Ochoa:2005ch}. These models have
the important feature of relating the number of quark families with the
number of colors, through the requirement of anomaly cancellation. As a
byproduct, the extension of the gauge group implies the presence of a new
neutral gauge boson $Z'$, which in general gives rise to flavor changing
neutral currents (FCNC) at the tree level. In addition, 331 models show
other interesting aspects, such as the presence of neutrino masses,
neutral and charged scalars, exotic quarks, etc.~which can be investigated
in the next generation of colliders like the LHC and the ILC. For example,
the nonstandard neutral current could be identified at the LHC by looking
at the process $pp\to Z'\to e^+ e^-$: performing specific kinematic cuts
on the outgoing electrons, it would be possible to reduce background so as
to distinguish the $Z'$ current within the 331 model from other theories
that include physics beyond the SM~\cite{Dittmar:2003ir}.

Concerning the presence of FCNC, it is important to point out that in 331
models it is not possible to fit all quark families in multiplets having
the same quantum numbers. As a consequence, while the $Z$ couplings to
ordinary quarks and leptons remain the same as in the SM, the
corresponding $Z'$ couplings are not universal for all quark families.
This gives rise to tree level flavor violation when rotating from the
current quark basis to the mass eigenstate basis. The size of the
couplings depends on the angles and phases of the (left-handed) up and
down quark mixing matrices $V_L^u$ and $V_L^d$, which therefore become
separately observable (in the framework of the SM, only the elements of
the matrix $V_{CKM} = V^{u\dagger}_L V_L^d$ can be measured). Due to the
unitarity of these mixing matrices, predictions for FCNC observables in
the 331 models are in general related to each other. In order to establish
bounds for this new physics, the most interesting sector is that of the
down-like quarks $d$, $s$ and $b$, where there are several well-measured
observables at our disposal. Here we concentrate on FCNC processes in
which flavor changes by two units. These typically show the most important
suppression within the SM, and consequently the most stringent bounds for
new physics. We will restrict ourselves to the down quark sector, taking
into account the following experimental data for five $\Delta F=2$
observables, where $F=S,B$~\cite{Yao:2006px}:
\begin{eqnarray}
\Delta m_K & = & m_{K_L} - m_{K_S} \ = \ (5.292\pm 0.009)\times 10^{-3}\
{\rm ps}^{-1} \nonumber \\
\Delta m_d & = & m_{B_H^0} - m_{B_L^0} \ = \ 0.507\pm 0.005\ {\rm ps}^{-1}
\nonumber \\
\Delta m_s &  = & m_{B_{sH}^0} - m_{B_{sL}^0} \ = \ 17.77\pm 0.12\
{\rm ps}^{-1} \nonumber \\
|\varepsilon_K| & = & (2.232\pm 0.007) \times 10^{-3} \nonumber \\
\sin \Phi_d & = & 0.687 \pm 0.032
\label{exp}
\end{eqnarray}
Here $\epsilon_K$ and $\Phi_d$ are CP-violating parameters, defined in
connection with $K^0-\bar K^0$ and $B_d^0-\bar B_d^0$ mixing respectively
(in fact, $\Phi_d$ arises from the interference between CP violation in
$B_d$ mixing and decay, the latter usually assumed to be negligible). It
is important to stress that the measurement of $\Delta m_s$, recently
obtained~\cite{Abulencia:2006ze}, is the first accurate experimental value
of a $\Delta S=\Delta B=2$ observable, and has attracted significant
theoretical interest~\cite{dmslit,Cheung:2006tm}. As stated, the $Z'$
contribution to this quantity in the 331 model can be directly related
with the contributions to the other observables in Eq.~(\ref{exp}),
allowing us to perform a global fit of the allowed region for the
down-like quark mixing parameters. This represents the main motivation for
the present work.

In the literature there are different versions of the 331 models,
according to the fermion content and quantum numbers, and the number of
scalar $SU(3)_L$ multiplets needed to break the gauge symmetry so as to
provide fermion masses. In general, these theories also include exotic
quarks of nonstandard charges. The first versions of the models included
three scalar triplets and one scalar sextet~\cite{pleitez}, and new
``quarks'' with electric charges $5/3$ and $-4/3$ (in fact, exotic
fermions can carry both quark and lepton numbers different from zero). For
definiteness and simplicity we will consider here a particular 331 model
that has been called ``economical''~\cite{Ponce:2002sg}, since it deals
with a minimal scalar sector of only two triplets, and does not include
fermions with nonstandard charges, i.e., other than $2/3$, $-1/3$ for
``quarks'' and 0 or $-1$ for ``leptons''. Recently, the ability of this
model to reproduce the observed neutrino mass pattern has been
discussed~\cite{Dong:2006mt}, and a supersymmetric version of the model
has been presented~\cite{Dong:2007qc}.

The paper is organized as follows: in Sect.\ 2 we present an overview of
331 models, focusing on the $Z'$-mediated neutral currents. In Sect.\ 3 we
derive the expressions for the new contributions to $\Delta F=2$
observables. Our numerical analysis, including a comparison with the
expected results using a definite ansatz for quark mass matrices, is
presented in Sect.\ 4. Finally, in Sect.\ 5 we summarize our results.

\section{Nonuniversal couplings in economical 331 models}

As stated earlier, in 331 models the SM gauge group is enlarged to
$SU(3)_C\otimes SU(3)_L\otimes U(1)_X$. The fermions are organized into
$SU(3)_L$ multiplets, which include the standard quarks and leptons, as
well as exotic particles usually called $J_i$, $E_i$ and $N_i$. Though the
criterion of anomaly cancellation leads to some constraints in the fermion
quantum numbers, still an infinite number of 331 models is allowed. In
general, the electric charge can be written as a linear combination of the
diagonal generators of the group,
\begin{equation}
Q\ = \ T_3\,+\,\beta\, T_8\,+\,X\ ,
\end{equation}
where $\beta$ is a parameter that characterizes the specific 331 model
particle structure and quantum numbers.

The organization of the three fermion families in 331 models is sketched
in Table I, where $i$ labels the quark family in the interaction basis,
and $\alpha = e,\mu,\tau$. Notice that the charges of the exotic particles
depend on the chosen value of the parameter $\beta$. As stated in the
Introduction, the so called ``economical'' 331 models~\cite{Ponce:2002sg}
are defined as those that do not include fermions with nonstandard
charges. Given the structure in Table I, this is possible only if one
takes $\beta=\pm 1/\sqrt{3}$, plus and minus sign corresponding to exotic
leptons of charge $-1$ and 0 (the correspondence is convention dependent).
Concerning the scalar sector, in the economical models it is possible to
give masses to all fermions and to reproduce the desired symmetry breaking
pattern with only two scalar triplets, usually called $\chi$ and $\eta$.
Choosing $\beta = 1/\sqrt3$, the vacuum expectation values of these scalar
fields can be written as $\langle\chi\rangle = 1/\sqrt2(0,u,w)^T$ and
$\langle\eta\rangle = 1/\sqrt2(v,0,0)^T$, while for $\beta = -1/\sqrt3$
one has $\langle\chi\rangle = 1/\sqrt2(u,0,w)^T$ and $\langle\eta\rangle =
1/\sqrt2 (0,v,0)^T$. The spontaneous gauge symmetry breaking proceeds into
two steps: a first breaking $SU(3)_L\otimes U(1)_X\to SU(2)_L\otimes
U(1)_Y$ at the energy scale given by the VEV $w$, and a second SM-like
breaking at a scale $v \sim 250$~GeV. As usual, fermion masses are
obtained from Yukawa-like couplings with the scalar fields. It is seen
that the model is able to provide the observed fermion mass pattern, where
the VEV $w$ sets the mass scale for the exotic
fermions~\cite{Dong:2006gx}. Bounds for the $SU(3)_L\otimes U(1)_X$
breaking energy scale provide a lower value for $w$ in the TeV
range~\cite{zbounds}.

\begin{table}[htb]
\begin{center}
\begin{tabular}{cccc}
Fermion & Representation & Q & X \\ & &  \\ \hline  & & \\
$\left(\begin{array}{c} d_1 \\ u_1 \\ J_1 \end{array}\right)_L$\ , \
$\left(\begin{array}{c} d_2 \\ u_2 \\ J_2 \end{array}\right)_L$
& {\bf 3$^\ast$} &
$\left(\begin{array}{c} -\frac{1}{3} \\ \frac{2}{3} \\
\frac{1}{6}+\frac{\sqrt{3}\beta}{2} \end{array}\right)$ &
$-\frac{1}{6}-\frac{\beta}{2\sqrt{3}}$ \\ \\
$\left(\begin{array}{c} u_3 \\ d_3 \\ J_3 \end{array}\right)_L$
& {\bf 3} & $\left(\begin{array}{c} \frac{2}{3} \\ -\frac{1}{3} \\
\frac{1}{6}-\frac{\sqrt{3}\beta}{2} \end{array}\right)$ &
$\frac{1}{6}-\frac{\beta}{2\sqrt{3}}$ \\ \\
$u_{iR}$ & {\bf 1} & $\frac{2}{3}$ & $\frac{2}{3}$ \\ \\
$d_{iR}$ & {\bf 1} & $-\frac{1}{3}$ & $-\frac{1}{3}$  \\ \\
$J_{1R}$, $J_{2R}$ & {\bf 1} & $\frac{1}{6}+\frac{\sqrt{3}\beta}{2}$
& $\frac{1}{6}+\frac{\sqrt{3}\beta}{2}$ \\ \\
$J_{3R}$ & {\bf 1} & $\frac{1}{6}-\frac{\sqrt{3}\beta}{2}$
& $\frac{1}{6}-\frac{\sqrt{3}\beta}{2}$ \\ \\
$\left(\begin{array}{c} \nu_\alpha \\ l_\alpha \\ F_\alpha
\end{array}\right)_L$
& {\bf 3$^\ast$} & $\left(\begin{array}{c} \mbox{\small 0} \\
\mbox{\small $-$1} \\ -\frac{1}{2}-\frac{\sqrt{3}\beta}{2} \end{array}\right)$ &
$-\frac{1}{2}-\frac{\beta}{2\sqrt{3}}$ \\ \\
$l_{\alpha R}$ & {\bf 1} & \mbox{\small $-$1} & \mbox{\small $-$1} \\ \\
$F_{\alpha R}$ & {\bf 1} & $-\frac{1}{2}-\frac{\sqrt{3}\beta}{2}$
& $-\frac{1}{2}-\frac{\sqrt{3}\beta}{2}$ \\ \\
\hline
\end{tabular}
\end{center}
\caption{Fermion representations and quantum numbers in 331 models}
\end{table}

Due to the enlarged group structure of the 331 models, one finds three
neutral gauge bosons $W^3$, $W^8$ and $B$. It is convenient to rotate
these states into a new basis where one can identify the usual SM gauge
fields $A$ and $Z$, together with a new $Z'$ state. The corresponding
transformation for arbitrary $\beta$ reads
\begin{eqnarray}
A_\mu & = & S_W\, W_\mu^3+C_W(\beta\, T_W\, W\mu^8
+ \sqrt{1-\beta^2T_W^2}\, B_\mu) \nonumber \\
Z_\mu & = & C_W\, W_\mu^3-S_W(\beta\, T_W\, W\mu^8
+ \sqrt{1-\beta^2T_W^2}\, B_\mu) \nonumber \\
Z^\prime_\mu & = & -\sqrt{1-\beta^2T_W^2}\, W_\mu^8
+ \beta\, T_W\, B_\mu \ ,
\end{eqnarray}
where we have introduced a Weinberg angle ($S_W = \sin\theta_W$, etc.).
This angle can be written in terms of the coupling constants $g$ and $g'$,
corresponding to the $SU(3)_L$ and $U(1)_X$ groups, respectively, as
\begin{equation}
T_W \ = \ \frac{g^\prime}{\sqrt{g^2+\beta^2g^{\prime2}}} \ .
\end{equation}

With this definition of $\theta_W$ the couplings of $A$ and $Z$ bosons to
ordinary fermions are the standard ones. We are interested now in the
couplings of the new $Z'$ state to ordinary quarks, in particular, to
down-like quarks $d$, $s$ and $b$, since we will deal here with neutral
$K$, $B_d$ and $B_s$ mesons. In terms of the electroweak current
eigenstates $D = (d_1\ d_2\ d_3)^T$, it can be seen~\cite{Ochoa:2005ch}
that the corresponding interaction Lagrangian is given by
\begin{eqnarray}
{\cal L}^{\rm (d)}_{\rm NC} & = &
\frac{g^\prime}{2\,S_W\,C_W} \ \bigg[
\sum_{i=1}^2 \bar D_i\; \gamma_\mu
\bigg(\frac{C_W^2}{\sqrt{3}} \; -\; \beta\, Q_d\,S_W^2 \bigg)\, P_L \, D_i
\nonumber \\
& & + \ \bar D_3\; \gamma_\mu
\bigg(- \frac{C_W^2}{\sqrt{3}} \; -\; \beta\, Q_d\,S_W^2 \bigg)\, P_L \, D_3
\; + \; \sum_{i=1}^3 \bar D_i \; \gamma_\mu \;
2\,\beta \, Q_d \, S_W^2 \; P_R \, D_i \bigg] \ Z^{\prime\mu} \ ,
\label{dnc}
\end{eqnarray}
where $P_{L,R} = (1\mp\gamma_5)/2$ and $Q_d = -1/3$. An important feature
shown in Eq.~(\ref{dnc}) is the fact that $Z'$ couplings to left-handed
quarks are not flavor-diagonal. This is a consequence of the group
structure of the 331 models shown in Table I: the requirement of anomaly
cancellation is satisfied only if one of the quark families is in a
different $SU(3)_L$ representation than the other two, which leads to
different quark-$Z'$ couplings. On the other hand, it is worth to notice
that the $Z'$ couplings to right-handed quarks turn out to be flavor
diagonal. Moreover, notice that in the case of left-handed quarks the
nondiagonal part of the interaction depends on the choice of $\beta$ only
through the value of the global coupling constant $g'/2S_W C_W$. In terms
of $\theta_W$ and $g$, one has
\begin{equation}
\frac{g'}{2 S_W C_W} \ = \ \frac{g}{2 C_W \sqrt{1-(1+\beta^2)S_W^2}} \ .
\label{gprima}
\end{equation}
In this way, since phenomenologically the value of $S_W$ at the
electroweak breaking scale is close to 1/4, the choices
$\beta=\pm\sqrt{3}$ leads to an enhancement of the quark-$Z'$ couplings.
For example, the ratio between the couplings $g'$ in the economical
($\beta=\pm1/\sqrt{3}$) and original ($\beta=-\sqrt{3}$) versions of the
331 models at the $m_Z$ scale is given by
\begin{equation}
\frac{g'_{\beta=\pm 1/\sqrt3}}{g'_{\beta=\pm\sqrt3}} \ \; = \ \;
\bigg( \frac{1\, -\, 4\,S_W^2}{1\, -\, \frac{4}{3}\, S_W^2} \bigg)^{1/2}
\ \simeq \ 0.33 \ .
\label{gratio}
\end{equation}
In the particular case of economical 331 models, the couplings in
Eq.~(\ref{dnc}) can be written as
\begin{eqnarray}
{\cal L}^{\rm (d)}_{\rm NC} & = & - \;
\frac{g^\prime}{\sqrt{3}\,S_W\,C_W} \ \bigg[
\sum_{i=1}^3 \bar D_i\; \gamma_\mu\, \bigg(
\epsilon_L'^{\,d(\pm)} \, P_L \; + \;
\epsilon_R'^{\,d(\pm)} \, P_R
\bigg) \, D_i \; + \; \bar D_3\; \gamma_\mu
C_W^2 \, P_L \, D_3 \bigg]\; Z^{\prime\mu}\ ,
\label{dnceco}
\end{eqnarray}
where
\begin{equation}
\epsilon_L'^{\,d(\pm)} \ = \ - \frac{1}{2} + \frac{3\mp 1}{6}\,S^2_W \ , \qquad
\epsilon_R'^{\,d(\pm)} \ = \ \pm\frac{1}{3}\,S^2_W \ \ ,
\label{estandard}
\end{equation}
$+$ and $-$ signs in $(\pm)$ corresponding to $\beta = 1/\sqrt{3}$ and
$\beta = -1/\sqrt{3}$, respectively.

Finally, let us point out that in general the states $Z$ and $Z'$ are only
approximate mass eigenstates, while the true physical states $Z_1$ and
$Z_2$ can be obtained from the former after a rotation. The corresponding
mixing angle $\theta$ is expected to be small, since it becomes suppressed
by a factor $r^2 \sim (v/w)^2$, i.e., the square of the ratio
between the $SU(3)_L\otimes U(1)_X \to SU(2)_L \otimes U(1)_Y$ and
$SU(2)_L\otimes U(1)_Y\to U(1)_{\rm em}$ symmetry breaking scales. In the
case of economical 331 models, at leading order in $r$ one
finds~\cite{Ochoa:2005ch,Van Dong:2006dr}
\begin{equation}
\theta \ = \ \frac{\sqrt{3-4S_W^2}}{4C_W^4} \;
\frac{[v^2 + (2S_W^2-1) u^2]}{w^2} \ \ .
\end{equation}
Though this angle will be in general small, $Z-Z'$ mixing will induce
flavor changes. Thus, in principle, this mixing has to be taken into
account when looking for observable effects of FCNC [see
Eqs.~(\ref{cijkl}-\ref{zz}) below].

\section{Theoretical expressions for $\Delta F=2$ observables}

In order to derive the theoretical expressions for the neutral meson
mixing observables in the above introduced economical 331 model, we take
into account the general analysis carried out in
Ref.~\cite{Langacker:2000ju}, considering the 331 theory as a particular
case. Thus we write the neutral current Lagrangian as
\begin{equation}
{\cal L}_{\rm NC} \ = \ - \; e\, J_{em}^\mu\,A_\mu\; -
\; g_1\,J^{(1)\mu}\,Z_\mu\; - \; g_2\,J^{(2)\mu}\,Z'_\mu  \ ,
\end{equation}
where $g_1 = g/C_W$, and the currents associated with the $Z$ and $Z'$
gauge bosons are
\begin{eqnarray}
J^{(1)}_\mu & = & \sum_i \bar q_i\, \gamma_\mu\, (\epsilon^q_L
\, P_L + \epsilon^q_R \, P_R) q_j \ \ , \\
J^{(2)}_\mu & = & \sum_{ij} \bar q_i\, \gamma_\mu\, (G^q_{L\, ij}
\, P_L + G^q_{R\, ij} \, P_R) q_j\ \ .
\end{eqnarray}
As in the previous section, here the fermions $q_i$ as well as the gauge
bosons $Z$ and $Z'$ are assumed to be gauge eigenstates. We will restrict
again to the couplings involving the down-like quark sector, where
$\epsilon_{L,R}^d$ are given by $\epsilon_L^d = -\frac{1}{2}+\frac{1}{3}
S_W^2$, $\epsilon_R^d =\frac{1}{3} S_W^2$, whereas $G_{L,R}^d$ are in
general $3\times 3$ matrices.

Let us consider now the effective four-fermion interaction Lagrangian for
the down quark sector in the mass eigenstate basis $D_i$, with $D = (d\
s\ b)^T$. As stated in Ref.~\cite{Langacker:2000ju}, one has
\begin{equation}
{\cal L}_{\rm eff} \ = \ -\;\frac{4\,G_F}{\sqrt{2}} \sum_{ijkl} \sum_{XY}
\; C^{ijkl}_{XY}\ (\overline{D}_i\,\gamma^\mu\, P_X\,D_j)\
(\overline{D}_k\,\gamma_\mu\, P_Y\,D_l)\ ,
\end{equation}
where $X$ and $Y$ run over the chiralities $L,R$, and $i,j,k,l$ label the
quark families. Assuming a small $Z-Z'$ mixing angle $\theta$, the
coefficients $C^{ijkl}_{XY}$ are given by~\cite{Langacker:2000ju}
\begin{equation}
C^{ijkl}_{XY} \ = \ \rho_{\rm eff}\, \delta_{ij}\, \delta_{kl}\,
\epsilon^d_X \, \epsilon^d_Y \; +\; y\,\delta_{ij}\, \epsilon^d_X \,
B^{d}_{Y\,kl} \; + \; y\,\delta_{kl}\,\epsilon^d_Y \, B^{d}_{X\,ij}\; +\;
z\,\rho\,\bigg(\frac{g_2}{g_1}\bigg)^2 \,B^{d}_{X\,ij}\, B^{d}_{Y\,kl}\ ,
\label{cijkl}
\end{equation}
where
\begin{eqnarray}
\rho_{\rm eff} & \simeq & \rho \ = \ \frac{m_W^2}{m_Z^2 C^2_W} \\
y & \simeq & \frac{g_2}{g_1}\; \rho \; \sin\theta\; \cos\theta \\
z & = & (\sin^2\theta + \frac{m_Z^2}{m_{Z'}^2} \cos^2\theta) \ .
\label{zz}
\end{eqnarray}
The presence of flavor changing neutral currents arises from the
nondiagonal elements of the $3\times 3$ matrices $B^d_{L,R}$.
Denoting by $V^u_{L,R}$ and $V^d_{L,R}$ the transformation matrices that
diagonalize the mass matrices for up and down quarks, one has
\begin{equation}
B^d_X \ = \ V_X^{d\,\dagger}\;G_X^d\; V_X^d\ ,
\end{equation}
and the usual CKM quark mixing matrix is given by
\begin{equation}
V_{CKM} \ = \ V_L^{u\dagger} \;V_L^d \ .
\end{equation}

{}From these general expressions it is immediate to obtain the effective
interaction Lagrangian in the economical 331 models. For $\beta=\pm
1/\sqrt3$, and introducing the definition
\begin{equation}
g_2 \ = \ \frac{g'}{\sqrt{3}\,S_W C_W} \ ,
\end{equation}
{}from Eq.~(\ref{dnceco}) one has $G_R^{d}  = \epsilon_R'^{\,d(\pm)}
\; \openone_{3\times 3}$, $G_L^{d}  = \epsilon_L'^{\,d(\pm)} \;
\openone_{3\times 3} \, +\, {\rm diag}(0,0,\cos^2\theta_W)$, and
\begin{eqnarray}
B_R^{d} & = & \epsilon_R'^{\,d(\pm)} \; \openone_{3\times 3} \nonumber \\
B_L^{d} & = & \epsilon_L'^{\,d(\pm)} \; \openone_{3\times 3} \; +\;
\cos^2\theta_W\;V_L^{d\,\dagger}\;{\rm diag}(0,0,1)\; V_L^d\ .
\end{eqnarray}
The contribution of ${\cal L}_{\rm eff}$ to $\Delta S = 2$ and $\Delta B =
2$ processes is driven by the coefficients $C^{ijkl}_{XY}$ with $i\neq j$,
$k\neq l$, which are proportional to the nondiagonal elements of the
$B^d_{X,Y}$ matrices. Therefore, for the economical 331 model, the
corresponding effective interaction will be given by
\begin{equation}
{\cal L}_{\rm eff} \ = \ -\;\frac{4\,G_F}{\sqrt{2}}\,\rho\,
\bigg(\frac{g_2}{g_1}\bigg)^2 z \;
(\overline{D}_i\,\gamma^\mu\, P_L\,B_{L\,ij}^d\, D_j)\
(\overline{D}_k\,\gamma_\mu\, P_L\,B_{L\,kl}^d D_l)\ ,
\label{effective}
\end{equation}
with $i\neq j$, $k\neq l$. The nondiagonal elements of $B_L^d$ read
\begin{equation}
B_{L\,ij}^d \ = \ \cos^2\theta_W \; V^{d\,\ast}_{L\, 3i}\, V^d_{L\, 3j}\ ,
\end{equation}
whereas the coupling constant ratio can be written in terms of the Weinberg
angle as
\begin{equation}
\bigg(\frac{g_2}{g_1}\bigg)^2 \ = \ \frac{1}{3-4\sin^2\theta_W}\ .
\end{equation}

In order to deal with phases, one can write without loss of
generality~\cite{Branco:2004ya}
\begin{equation}
V_L^d \ = \ P \; \tilde V\; K
\label{vdef}
\end{equation}
where $P = {\rm diag}(e^{i\phi_1},1,e^{i\phi_3})$, $K={\rm
diag}(e^{i\alpha_1},e^{i\alpha_2},e^{i\alpha_3})$, while the unitary
matrix $\tilde V$ can be written in terms of three mixing angles
$\theta_{12}$, $\theta_{23}$ and $\theta_{13}$ and a phase $\varphi$ using
the standard parameterization~\cite{foot1}
\begin{equation}
\tilde V \ = \left(
\begin{array}{ccc}
c_{12}\,c_{13} & s_{12}\,c_{13} & s_{13}\,e^{-i\varphi} \\
-s_{12}\,c_{23}-c_{12}\,s_{23}\,s_{13}\,e^{i\varphi} &
c_{12}\,c_{23}-s_{12}\,s_{23}\,s_{13}\,e^{i\varphi} &
s_{23}\,c_{13} \\
s_{12}\,s_{23}-c_{12}\,c_{23}\,s_{13}\,e^{i\varphi} &
-c_{12}\,s_{23}-s_{12}\,c_{23}\,s_{13}\,e^{i\varphi} &
c_{23}\,c_{13}
\end{array}
\right)\ \ .
\label{vparam}
\end{equation}

Let us proceed to write down the theoretical expressions for the $\Delta
F=2$ observables under consideration. In general, they will receive both
SM contributions arising from standard one loop diagrams, together with
the new 331 contributions from tree level FCNC. Denoting by $M^{P}_{12}$
the matrix element $\langle P^0|{\cal H}_{\rm eff}|\bar P^0\rangle$, one
obtains
\begin{eqnarray}
\Delta m_K & = & 2\,{\rm Re}(M_{12}^{K(SM)}+M_{12}^{K(331)}) \label{uno} \\
\Delta m_d & = & 2\,\left|M_{12}^{B_d(SM)}+M_{12}^{B_d(331)}\right| \\
\Delta m_s & = & 2\,\left|M_{12}^{B_s(SM)}+M_{12}^{B_s(331)}\right| \\
\varepsilon_K & = & \frac{e^{i\pi/4}}{\sqrt{2} \Delta m_K}\;
{\rm Im}(M_{12}^{K(SM)}+M_{12}^{K(331)}) \\
\Phi_d & = & -\,{\rm arg}(M_{12}^{B_d(SM)}+M_{12}^{B_d(331)}) \ .
\label{cinco}
\end{eqnarray}

The corresponding SM contributions are well known~\cite{Buchalla:1995vs}.
One has
\begin{eqnarray}
M_{12}^{K(SM)} & = & \frac{G_F^2}{12\pi^2}\, m_W^2\, m_K\, f_K^2\,
\hat B_K \left[ \eta_1 S_0(x_c)\,\lambda_c^2\, +
\, \eta_2\, S_0(x_t)\,\lambda_t^2 \,+ \, 2\eta_3\,\lambda_c\,\lambda_t
\,S(x_c,x_t)\right] \label{dmk} \\
M_{12}^{B_q(SM)} & = & \frac{G_F^2}{12\pi^2}\, m_W^2\, m_{B_q}\,
f_{B_q}^2\, \hat B_{B_q} \eta_B \, S_0(x_t)\, (V_{tq} V_{tb}^\ast)^2\ ,
\label{dmbq}
\end{eqnarray}
where $S_0(x_q)$ are Inami Lim functions~\cite{inamilim} arising from box
diagram contributions, and $\hat B_P$, $\eta_i$, $\eta_B$ are parameters
that account for theoretical uncertainties related with both long- and
short-distance QCD corrections.

On the other hand, from the effective interaction in Eq.~(\ref{effective})
it is easy to obtain the relevant expressions for the 331 contributions.
These are given by
\begin{equation}
M_{12}^{P(331)} \ = \ \frac{2\sqrt{2}}{3}\,G_F\,\rho\, m_P\, f_P^2\,
\hat B_P\, \bigg(\frac{g_2}{g_1}\bigg)^2 \cos^4\theta_W \, z\;
\lambda_P^2\ ,
\label{zp}
\end{equation}
where
\begin{eqnarray}
\lambda_K & = & s_{13}\,s_{23}\,c_{13}\,e^{i(\phi_1-\varphi)}
\label{k} \\
\lambda_{B_d} & = & s_{13}\,c_{23}\,c_{13}\,e^{i(\phi_1-\phi_3-\varphi)}
\label{bd} \\
\lambda_{B_s} & = & s_{23}\,c_{23}\,c_{13}^2\,e^{-i\phi_3} \ \ .
\label{bs}
\end{eqnarray}
Thus, it is seen that the 331 contributions to the five observables in
Eqs.~(\ref{uno}-\ref{cinco}) are given in terms of five unknown
parameters, namely the suppression factor $z$ defined in Eq.~(\ref{zz}),
the angles $\theta_{13}$, $\theta_{23}$ and two CP-violating phases
coming from the $V_L^d$ mixing matrix. We choose here as independent
parameters the phases $\phi'\equiv \phi_1-\phi_3-\varphi$ and
$\phi''\equiv\phi_1-\varphi$, the remaining phase in Eq.~(\ref{bs}) being
$\phi_3 =\phi''-\phi'$.

\section{Inputs, numerical procedure and results}

As stated, our aim is to take into account the present experimental data
for the above mentioned $\Delta F=2$ observables in order to constrain the
values of the 331 parameters. Clearly, in order to perform this analysis
it is necessary to take into account both the theoretical and experimental
uncertainties in the determination of the respective SM contributions.

In our analysis, the experimental values of particle masses in
Eqs.~(\ref{uno}-\ref{dmbq}), as well as the kaon decay constant and the
value of $\sin \theta_W$ at the electroweak breaking scale have been taken
from the PDG Review~\cite{Yao:2006px}, while for the quark masses entering
the SM box diagrams we have used $m_c=1.3\pm 0.1$~GeV and $m_t =
168.5$~GeV. The theoretical estimations for the short-distance QCD
corrections $\eta_i$ and $\eta_B$ in Eqs.~(\ref{dmk}) and (\ref{dmbq})
have been taken as $\eta_1 = 1.32\pm 0.32$, $\eta_2 = 0.57\pm 0.11$,
$\eta_3 = 0.47\pm 0.05$ and $\eta_B = 0.55$~\cite{Buras:2005xt}. For the
value of the parameter $B_K$ we have used the recent lattice result $B_K =
0.83\pm 0.18$~\cite{Gamiz:2006sq}, while the values of the parameters
$B_{B_d}$ and $B_{B_s}$, as well as the $B_d$ and $B_s$ decay constants,
have been obtained by averaging results of unquenched lattice
calculations~\cite{Aoki:2003xb,Dalgic:2006gp}. This leads to $f_{B_d}
\sqrt{B_{B_d}} = 0.21 \pm 0.03$, $f_{B_s}\sqrt{B_{B_s}} = 0.25 \pm 0.03$.

Now, special care has to be taken when dealing with the parameters of the
CKM quark mixing matrix. The reason is that present global fits are
strongly dependent on theoretical results based on one loop SM processes,
which could be modified by the effect of 331 contributions. In this sense,
our procedure is similar to that in Ref.~\cite{Promberger:2007py}: instead
of using full CKM angle fits, we just take into account the experimental
constraints obtained from tree-level dominated processes. Thus, from the
Particle Data Group analysis we take~\cite{Yao:2006px}
\begin{eqnarray}
|V_{ud}| = 0.9738 \pm 0.0003 \qquad\qquad
|V_{us}| = 0.226 \pm 0.002 \qquad\qquad
|V_{ub}| = 0.0043 \pm 0.0003 & & \label{vu} \nonumber \\
|V_{cd}| = 0.230 \pm 0.011 \ \ \ \; \qquad\qquad
|V_{cs}| = 0.957 \pm 0.095 \qquad\qquad
|V_{cb}| = 0.0416 \pm 0.0006 & &
\end{eqnarray}
Then, as a further experimental input we take into account the value of the
CP-violating parameter $\gamma = {\rm arg}(-
V_{ud}V_{ub}^\ast/V_{cd}V_{cb}^\ast)$ obtained from tree-level dominated
$B\to D^{(*)}X$ decays. From the analyses carried out by
CKMfitter~\cite{Charles:2004jd} and UTfit~\cite{Bona:2005vz}
collaborations we get
\begin{equation}
\gamma \ = \ 78^\circ \pm 17^\circ \ .
\label{gamma}
\end{equation}

Taking into account this set of experimental values, we proceed to
estimate the allowed range for the 331 model parameters appearing in
Eqs.~(\ref{uno}-\ref{cinco}) compatible with the experimental measurements
of the five observables of interest. The $V_{CKM}$ matrix parameters are
treated as follows: in order to decide the compatibility of a given set of
331 parameter values, we consider a manifestly unitary parameterization of
the $V_{CKM}$ matrix [as that in Eq.~(\ref{vparam})], and let the values
of the mixing angles and the complex phase vary freely. The 331 parameter
set is kept only if the experimental constraints (\ref{exp}) are satisfied
and at the same time the corresponding set of $V_{CKM}$ parameters is
found to be compatible with the ranges in Eqs.~(\ref{vu}-\ref{gamma}). In
this way, we take care of the correlations between the error bars in the
331 parameters and the error bars in the experimental constraints on
$V_{CKM}$ arising from tree level dominated processes. Constraints on
$V_{CKM}$ elements involving the top quark as well as the CP-violating
angle $\beta$ will arise directly from the experimental values of $\Delta
F=2$ observables and the unitarity of the $V_{CKM}$ matrix in presence of
the 331 contributions.

\begin{figure}[htb]
\vspace*{0.5cm}
\centerline{
   \includegraphics[height=5.9truecm]{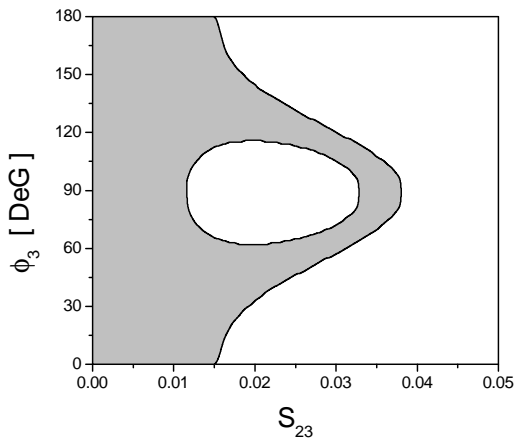}
   } \caption{Allowed $s_{23}-\phi_3$ region for $s_{13}=0$, $z =
(m_Z/1~{\rm TeV})^2$, considering a $2\sigma$ confidence level in the
experimental errors of $\Delta F=2$ observables.}
\label{fig:1}
\end{figure}
\begin{figure}[htb]
\vspace*{0.5cm}
\centerline{
   \includegraphics[height=5.9truecm]{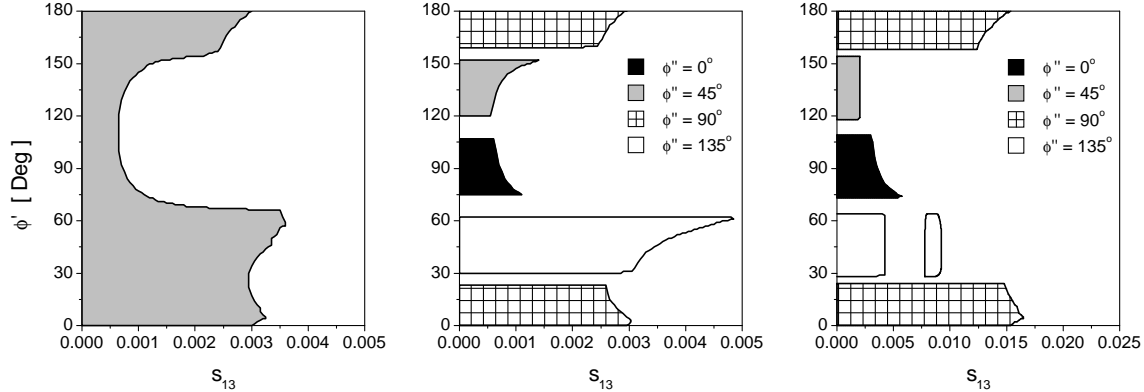}
   } \caption{Allowed $s_{13}-\phi'$ regions considering a $2\sigma$
confidence level in the experimental errors of $\Delta F=2$ observables.
Left panel corresponds to $z=z_1$ and low values of $s_{23}$, whereas
central and right panels correspond to $z = z_1$, $s_{23} = 0.036$ and $z
= z_2$, $s_{23} = 0.18$, respectively.}
\label{fig:2}
\end{figure}

Let us turn now to present our results. We begin by considering the 331
parameter region by demanding compatibility with the experimental values
(\ref{exp}) at the level of $2\sigma$. At this level the data can be
reproduced by the SM alone (i.e.~the values $\theta_{13} = \theta_{23} =
0$ lie within the allowed range). In order to deal with the five-parameter
space, let us first fix the value of $z$ as $z_1 = (m_Z/1~{\rm TeV})^2$,
and take the mixing angle $\theta_{13}=0$. In this case the only
constraint arises from Eq.~(\ref{bs}), which determines a region for
$s_{23}$ and $\phi_3 = \phi''-\phi'$. This is represented in Fig.~1, where
it is found that there is an upper bound $|s_{23}|\leq
0.038$~\cite{foot2}. We will not consider here the other possible
solution, $|s_{23}|\simeq 1$, $|c_{23}|\leq 0.038$, following the common
belief that assumes a correlation between the hierarchies in quark masses
and mixing angles. Then we consider the case $\theta_{23}=0$, in which the
constraint arises from Eq.~(\ref{bd}), and one finds an allowed region in
the $s_{13}$ and $\phi'$ plane, as shown in the left panel of Fig.~2. We
see here that the value of $|s_{13}|$ can be as large as 0.0035, depending
on the value of the phase $\phi'$. Considering nonzero values of $s_{23}$,
it is seen that this region remains unchanged if $s_{23}$ is relatively
low, while it becomes reduced when $s_{23}$ approaches the upper bound of
0.038. Close to this bound, only certain ranges for the phase $\phi'$ are
allowed, depending on the value of $\phi''$. This is shown in the central
panel of Fig.~2, where we have taken $s_{23}=0.036$ and some
representative values of $\phi''$. Let us now consider the dependence on
the $SU(3)_L$ symmetry breaking scale, increasing the value of $z$ from
$z_1$ to $z_2 = (m_Z/5~{\rm TeV})^2$. As expected, for low values of
$s_{23}$ the bounds for $s_{13}$ are just increased by a factor five, and
the same happens with the upper bound for $s_{23}$. In the right panel of
Fig.~2 we show the allowed regions in the $s_{13}-\phi'$, taking now
$s_{23}=5 \times 0.036 = 0.18$. While the ranges for $\phi'$ are
approximately the same as in the central panel for $s_{13}=0$, the
combined effects of all five experimental constraints produce some
distortions for larger values of $s_{13}$.

\begin{figure}[htb]
\vspace*{0.5cm}
\centerline{
   \includegraphics[height=5.7truecm]{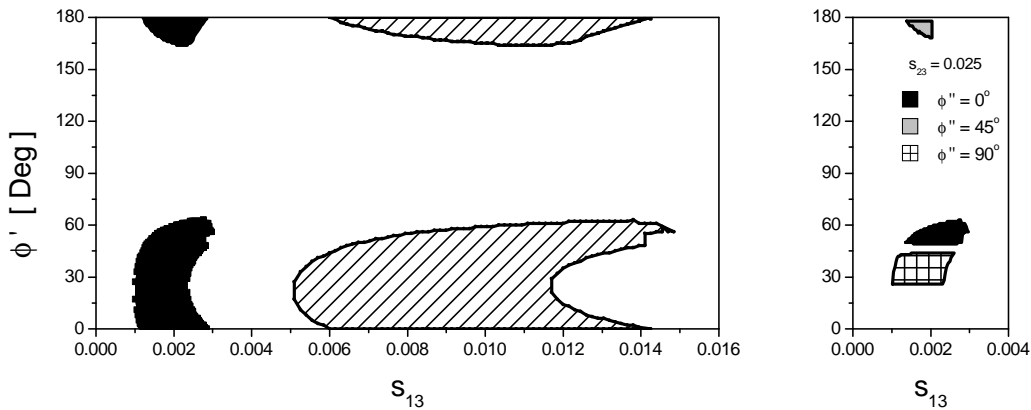}
   } \caption{Allowed $s_{13}-\phi'$ region for low values of $s_{23}$
(left) and $s_{23} = 0.025$ (right). In the left panel, black and shadowed
regions correspond to $z = (m_Z/1~{\rm TeV})^2$ and $z = (m_Z/5~{\rm
TeV})^2$, respectively. The regions in the right panel correspond to $z =
(m_Z/1~{\rm TeV})^2$ and different values of $\phi''$.}
\label{fig:3}
\end{figure}

Finally, we present the results corresponding to a confidence level of
1$\sigma$ in the experimental data. Once again, for low values of
$\theta_{23}$ the parameter range allowed for $\theta_{13}$ and $\phi'$ is
independent of $\theta_{23}$ and $\phi''$. The results are shown in the
left panel of Fig.~3, where black and shadowed areas correspond to $z=z_1$
and $z=z_2$ respectively. We find now that for $z=z_1$ the value of
$\theta_{23}$ is constrained by $|s_{23}|\leq 0.027$. As in the previous
case, close to this upper bound for $\theta_{23}$ the allowed regions
appear to be further constrained, and the corresponding reduced zones
depend on the value of $\phi''$. This is shown in the right panel of the
figure, where we show the allowed $s_{13} - \phi'$ parameter space for
$z=z_1$ taking now $s_{23} = 0.025$ and $\phi''=0$, 45, 90 and 135 degrees
(the latter leads to no solution). It is seen from this analysis that the
values $s_{13} = s_{23} = 0$ are not allowed, which means that the SM is
not able to reproduce the full set of experimental data at the level of
one standard deviation, requiring the presence of some new physics.

It is worth noting that our analysis can be also applied to other versions
of the 331 model, differing in the choice of the parameter $\beta$. Though
these versions would present different scalar and fermion quantum numbers,
the effect of this change on FCNC's driven by the $Z'$ boson can be
trivially taken into account. Indeed, as stated in Sect.\ 2, from
Eq.~(\ref{dnc}) it is seen that the change of $\beta$ affects the
nondiagonal part of $Z'$ currents just by rescaling the value of the
coupling $g'$. In the general case, one has for the 331 contributions to
$M_{12}^P$ [see Eq.~(\ref{zp})]
\begin{equation}
\left(\frac{g_2}{g_1}\right)^2 z\; \lambda_P^2 \ \sim \
\frac13\;\frac{1}{1-(1+\beta^2)S_W^2}\; \left(\frac{m_Z}{m_{Z'}}\right)^2
\;|(V^d_L)_{i3} (V^{d\ast}_L)_{j3}|^2\ ,
\end{equation}
thus one could take $(g_2/g_1)^2 z$ as the relevant 331 model parameter.
In this way it is possible to complement our results with those obtained
in Ref.~\cite{Promberger:2007py}, where the authors consider the effect of
FCNC's in the original version of the 331 model (i.e.~taking
$\beta=-\sqrt{3}$, within our sign conventions). Since in this model the
ratio $g_2/g_1$ is approximately enhanced by a factor of three [see
Eq.~(\ref{gratio})], we should reproduce their results just by scaling the
value of $z$ by a factor $\simeq 9$. Our results for $z=z_1$ would
correspond to those obtained in Ref.~\cite{Promberger:2007py} for
$m_{Z'}=3$~TeV (notice that in Ref.~\cite{Promberger:2007py} the authors
consider only some particular values for the angles $\phi'$ and $\phi''$,
and the $Z-Z'$ mixing angle is neglected). Indeed, considering only the
constraints imposed by the experimental values of $\Delta m_K$ and
$\varepsilon_K$, in this way we find good agreement with the results
obtained in Ref.~\cite{Promberger:2007py} for the bounds on $s_{13}$ and
$s_{23}$. In our paper the results are presented in a different way, since
we are considering the correlation between all five experimental
constraints in Eq.~(\ref{exp}).

\hfill

To conclude, let us analyze qualitatively the bounds obtained for $s_{13}$
and $s_{23}$. We recall that the down-like quark mixing angles
$\theta_{13}$ and $\theta_{23}$ are hidden parameters in the SM, where the
only observable quantities are the entries in the $V_{CKM}$ matrix. In
order to get some insight on the expected sizes of these mixing angles, it
is interesting to consider the values of $s_{13}$ and $s_{23}$ arising
from a definite ansatz for the mixing matrix $V_L^d$. For the sake of
illustration, we consider here the case of Hermitian quark mass matrices
having a four-zero texture~\cite{rev}. This is a simple and widely studied
ansatz, in which the down quark mass matrix has the
form~\cite{Fritzsch:2002ga}
\begin{equation}
\tilde M_d \ = \left(
\begin{array}{ccc}
0 & C_d & 0 \\
C_d^\ast & \tilde{B_d} & B_d \\
0 & B_d^\ast & A_d
\end{array}
\right)\ \ ,
\end{equation}
where (owing to the quark mass hierarchy $m_d \ll m_s \ll m_b$) one
expects $|A_d|\gg |\tilde{B_d}|$, $|B_d|$, $|C_d|$. The mixing matrix
$V_L^d$ can be written in terms of the quark masses and some additional
parameters. In particular, the matrix elements $\tilde V_{13}$ and $\tilde
V_{23}$, defined according to Eqs.~(\ref{vdef}-\ref{vparam}), are
approximately given by~\cite{Fritzsch:2002ga,Matsuda:2006xa}
\begin{equation}
|\tilde V_{13}| \ \simeq \
\sqrt{\frac{m_d\, m_s\, (m_b-A_d)}{A_d\; m_b^2}} \ \ ,
\qquad\qquad
|\tilde V_{23}| \ \simeq \ \sqrt{\frac{m_b-A_d}{m_b}} \ \ .
\end{equation}
where the value of $A_d/m_b$ is constrained by the experimental value of
the ratio of $V_{CKM}$ elements $|V_{ub}/V_{cb}|$. From this constraint
one obtains $0.88 \lesssim A_d/m_b \lesssim 0.98$~\cite{Matsuda:2006xa}.
Noting that $|s_{13}|\simeq |\tilde V_{13}|$ and $|s_{23}|\simeq |\tilde
V_{23}|$, one obtains
\begin{equation}
0.001 \lesssim |s_{13}| \lesssim 0.003 \ \ ,
\qquad\qquad
0.15 \lesssim |s_{23}| \lesssim 0.35 \ \ .
\label{ss}
\end{equation}
If one compares these bounds with the constraints obtained in the
framework of the economical 331 models from the experimental values of
$\Delta F=2$ observables (\ref{exp}), one achieves consistency with the
bounds for $|s_{23}|$ only if new physics shows up at a scale larger than
a few TeV, namely $z \lesssim (m_Z/\mbox{5 TeV})^2$. In this case the
range of $s_{13}$ in (\ref{ss}) would be somewhat low to reproduce the
experimental values in Eq.~(\ref{exp}) at the level of one standard
deviation (see Fig.~3), and consistency would be obtained at the $2\sigma$
level (see right panel of Fig.~2). In the case of the original version of
the 331 model, according to the previous discussion the bound for the new
scale should be extended to about 15 TeV.

\section{Summary}

We have analyzed here tree-level flavor changing neutral currents in the
context of economical 331 models, in particular, considering the
phenomenological bounds on model parameters arising from experimental
values of $\Delta F=2$ observables. In general, 331 models include the
presence of exotic fermions and gauge bosons, which could be observed in
forthcoming experiments such as LHC and ILC. At lower energies, one of the
most stringent tests for the model is provided by the effect of FCNC's,
which arise at tree level owing to the presence of nonuniversal couplings
of a neutral gauge boson $Z'$. Here we have concentrated on the study of
flavor mixing in the down quark sector, where $\Delta F=2$ observables
provide a set of experimental data that allows one to obtain the bounds
for the relevant model parameters.

Our parameter space includes five variables, namely the angles
$\theta_{13}$, $\theta_{23}$ and the CP-violating phases $\phi'$,
$\phi''$, coming from the $V_L^d$ mixing matrix, and the scale parameter
$z$ [or, in general, the combination $(g_2/g_1)\, z$]. In the economical
model, taking $z = (m_Z/{\rm 1\ TeV})^2$, we have found upper bounds for
the mixing angles $|\theta_{13}|\lesssim 0.003$ and $|\theta_{12}|\lesssim
0.035$. These bounds are in fact correlated, and depend on the values of
the phases $\phi'$ and $\phi''$. The allowed region for $\theta_{23}$ with
$\theta_{13}=0$ is shown in Fig.~1, while the allowed regions for
$\theta_{13}$ taking extreme values of $\theta_{23}$ are shown in Figs.~2
and 3 ($2\sigma$ and $\sigma$ confidence level, respectively). In general,
these last regions are found to depend both on $\phi'$ and $\phi''$. We
have also shown how these bounds scale with the value of $z$ increasing
the $SU(3)_L\otimes U(1)_X$ breaking scale by a factor five. Finally, for
the sake of illustration we have compared these results with the expected
values of the down quark mixing angles within a four-zero texture ansatz
for the mass matrices. We have found that imposing such an ansatz in the
context of economical 331 models would be compatible with experimental
data for FCNC observables at $2\sigma$ level, provided that the
$SU(3)_L\otimes U(1)_X$ symmetry breaking occurs at a scale above 5 TeV.

\section{Acknowledgments}

This work has been supported in part by CONICET and ANPCyT (Argentina,
grants PIP 6009 and PICT04-03-25374), Fundaci\'on Banco de la Rep\'ublica
(Colombia), and the High Energy Physics Latin-American-European Network
(HELEN).


\begin{thebibliography}{50}
\bibitem{foot1} In fact the unitarity of $\tilde V$ is only approximate,
owing to the possible mixing between exotic and ordinary quarks. We work
in the limit where exotic quarks decouple, which is justified by the large
value of the scale $w$ in comparison with the bottom quark mass.
\end{thebibliography}
\end{document}